\documentclass{Interspeech}

\interspeechcameraready

\title{Amplifying Artifacts with Speech Enhancement in Voice Anti-spoofing}
\author[affiliation={1}, equalcontribution]{Thanapat}{Trachu}
\author[affiliation={2}, equalcontribution, daggercontribution]{Thanathai}{Lertpetchpun}
\author[affiliation={1}]{Ekapol}{Chuangsuwanich}

\affiliation{Department of Computer Engineering, Faculty of Engineering}{Chulalongkorn University}{Thailand}
\affiliation{University of Southern California}{Los Angeles}{USA}
\email{thanapat.trachu@gmail.com, lertpetc@usc.edu, ekapol.c@chula.ac.th}

\usepackage{comment}

\begin{document}

\maketitle

% Define footnotes manually
% \renewcommand{\thefootnote}{\fnsymbol{footnote}}
% \footnotetext[1]{These authors contributed equally.}
% \footnotetext[2]{Work performed at Chulalongkorn University.}
% \renewcommand{\thefootnote}{\arabic{footnote}} % Restore default if needed

% \def\thefootnote{*}\footnotetext{}
% \def\thefootnote{\dagger}\footnotetext{Work performed at Chulalongkorn University}

\begin{abstract}
Spoofed utterances always contain artifacts introduced by generative models. While several countermeasures have been proposed to detect spoofed utterances, most primarily focus on architectural improvements. In this work, we investigate how artifacts remain hidden in spoofed speech and how to enhance their presence. We propose a model-agnostic pipeline that amplifies artifacts using speech enhancement and various types of noise. Our approach consists of three key steps: noise addition, noise extraction, and noise amplification. First, we introduce noise into the raw speech. Then, we apply speech enhancement to extract the entangled noise and artifacts. Finally, we amplify these extracted features. Moreover, our pipeline is compatible with different speech enhancement models and countermeasure architectures. Our method improves spoof detection performance by up to 44.44\% on ASVspoof2019 and 26.34\% on ASVspoof2021.
% Furthermore, our analysis show that spoofeding artifacts dominantly lies in lower frequency bands.

% 129 words 920 characters
\end{abstract}

\section{Introduction}
% voice anti spoofeding in general LA, PA, deepfake, SASV and say we only interested in LA

The voice anti-spoofing task involves classifying whether an input utterance is spoofed or bona fide. To protect automatic speaker verification (ASV) systems from spoofing attacks, a separate system called a countermeasure (CM) is typically developed to detect spoofed utterances. In the ASVspoof2019 challenge \cite{asvspoof2019}, two main tasks were introduced: physical attacks (PA) and logical attacks (LA). PA focuses on defending against replay attacks, while LA targets speech generated by text-to-speech (TTS) and voice conversion (VC) systems. The ASVspoof2021 challenge \cite{asvspoof2021} introduced codecs to increase the difficulty of defending against attacks. \cite{10446798} Deng et al. proposed a pipeline to detect spoofed voices using vocoder fingerprints, indicating that different VCs have different artifacts. In ASVspoof2024 \cite{asvspoof2024}, adversarial attacks were included alongside new TTS and VC architectures to ensure countermeasures remain robust against evolving spoofing techniques. In this work, we focus on logical attacks, where utterances are generated using text-to-speech and voice conversion models.

% Beyond these challenges, there are other variations of the anti-spoofing task, such as Partial Anti-Spoofing \cite{partial_anti_spoofing}, where the model predicts which part of the utterance is spoofed, and Spoofing-Aware Speaker Verification (SASV) \cite{jung2022sasv}, which integrates countermeasure and ASV systems into a single framework. 

% % talk about artifact on VC and TTS
% It it generally known that generative models always generate artifacts. Text to speech (TTS) and voice conversion (VC) is no exception. They 

% speech enhancement in general
Speech enhancement aims to remove noise from a noisy speech signal to produce a cleaner output. It is commonly used alongside other speech processing tasks, such as speech recognition \cite{10122599, maas12_interspeech, koizumi22b_interspeech} and speaker verification \cite{shon19b_interspeech}, to enhance system robustness against noise.
However, as shown in \cite{storm_se}, speech enhancement does not solely extract noise but may also remove unrelated information, including parts of the clean speech signal. Given this behavior, we expect that applying speech enhancement to spoofed utterances could similarly remove artifacts.
% In general, the objective functions of speech enhancement include mean square error (MSE) between the model's output and the clean speech \cite{gagnet, trachu24_interspeech}, or the signal-to-noise ratio in the model's output \cite{convtasnet}.

% our intuition
% Previous work
Wang et al. \cite{wang23v_interspeech} used speech enhancement for noise reduction in anti-spoofing. In contrast, we leverage speech enhancement as an artifact enhancer to improve countermeasure performance against spoofing attacks. Our approach involves adding noise to the input, extracting the entangled noise and artifacts using speech enhancement, and then amplifying both before reinserting them into the utterance. As a result, spoofed speech exhibits a higher artifact magnitude, while bona fide speech behaves like typical noisy speech.

% result
The proposed pipeline was evaluated on the ASVspoof2019 and ASVspoof2021 datasets. Our results show that it improves countermeasure performance by 44.44\% relative improvement and remains robust across different speech enhancement and countermeasure architectures. Additionally, our findings suggest that artifacts located in lower frequency bands are more important than that in higher frequency bands for detecting spoof.

% \section{Related Work}
% Some studies have explored methods to amplify features to improve model performance. \cite{8756541}, for example, introduces a step that enhances subtle facial movements, making them easier for the classification model to detect. To the best of our knowledge, the concept of feature amplification in the input domain to improve detection remains underexplored in the voice anti-spoofing field. This motivates us to propose a step that amplifies artifacts using a speech enhancement model.

\section{Background on CM and SE}
\subsection{Countermeasures}
Countermeasure (CM) systems are designed to determine whether an input utterance is bona fide or spoofed. Various approaches exist for developing countermeasures, including using mel spectrograms with convolutional neural networks, input waveforms with graph neural networks, or end-to-end architectures. Light Convolutional Neural Network (LCNN) \cite{lavrentyeva19_interspeech} was the baseline for ASVspoof2021 and supports multiple feature inputs, such as LFCC, CQT, FFT, and mel spectrograms. The architecture consists of a long sequence of convolutional layers, Max-Feature-Map (MFM) activation, and batch normalization. AASIST \cite{aasist} and RawNet2 \cite{rawnet2} were introduced as baselines for ASVspoof2024. AASIST was designed to extract spectro-temporal features from the waveform using a Heterogeneous Stacking Graph Attention Layer (HS-GAL) and Max Graph Operations (MGO). RawNet2 employs Sinc filters to transform the waveform into features for the model and primarily consists of residual blocks (batch normalization, leaky ReLU, convolutional layers, pooling layers, and feature map scaling), a Gated Recurrent Unit (GRU), and a classifier. While these models differ in architecture and input features, they achieve comparable performance. In our work, we adopt LCNN, AASIST, and RawNet2 as the main baselines.

\subsection{Speech Enhancement Model}
% Overall general how to train SE and the objective
Speech enhancement aims to improve the quality and intelligibility of speech by modeling noisy speech as the sum of clean speech and noise. There are various approaches to developing speech enhancement, including generative \cite{trachu24_interspeech, sgmse} and discriminative \cite{gagnet, convtasnet, ncsnpp} methods.

Thunder \cite{trachu24_interspeech}, for example, is a generative model based on the diffusion process, which consists of two stages: the forward process and the reverse process. In the forward process, noise is gradually added to the data until it transforms into a predefined distribution, known as the prior distribution. The reverse process then removes noise from a sample drawn from the prior distribution, eventually restoring the clean speech. To perform the reverse process, a model is required to predict the noise in the input, iteratively removing it to recover the clean speech. This type of model is typically trained using mean squared error (MSE) between the output and the added noise.
NCSNPP \cite{ncsnpp}, on the other hand, is a discriminative model optimized with a different objective function: the signal-to-noise ratio (SNR). While both Thunder and NCSNPP are based on U-Net architectures, GaGNet \cite{gagnet} introduces a new architecture that mimics the human ability to focus on both coarse and fine-grained details. GaGNet decomposes speech enhancement into two sub-tasks: glance and gaze. The glance path suppresses noise in the magnitude domain of the spectrogram, while the gaze path refines the spectrogram in the complex domain. The model is then trained using the MSE loss function in the frequency domain.
In our work, we use Thunder, NCSNPP, and GaGNet to demonstrate that our method improves countermeasure performance regardless of the choice of the speech enhancement model.

\section{Method}
Our proposed approach leverages an off-the-shelf speech enhancement model to amplify artifacts in spoofed utterances. We hypothesize that while speech enhancement models are designed to remove noise, they also eliminate spoofing artifacts present in audio. By extracting and amplifying these artifacts, we aim to make them more distinguishable for the countermeasure system.

\begin{figure}[t] %ht
    \centering
    \includegraphics[scale=0.7]{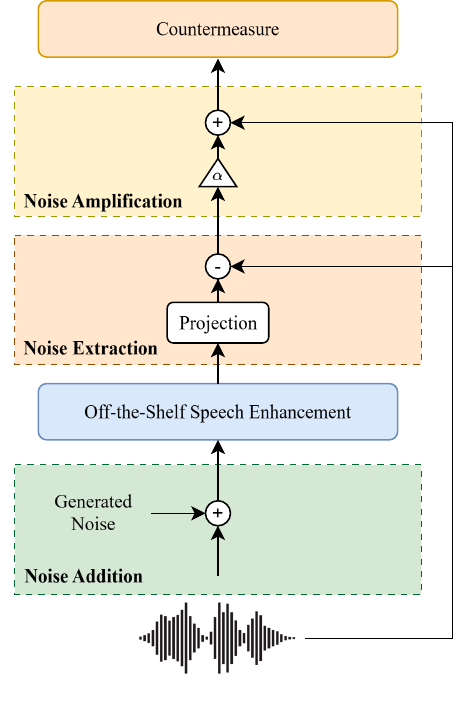}
    \caption{\textbf{Overview of our method.} First, noise is added to the input utterances, which are then processed by the speech enhancement model. Noise and artifacts are extracted from the output and reintroduced into the input utterances.}
    \vspace{-5mm}
    \label{fig:pipeline}
\end{figure}

An overview of our pipeline is presented in Figure \ref{fig:pipeline}. The pipeline consists of three key steps: (1) noise addition, (2) noise extraction, and (3) noise amplification. Finally, the countermeasure model is trained using the amplified utterances. In this section, we provide a detailed explanation of each step.

% An overview of our pipeline is presented in Figure \ref{fig:pipeline}. Our pipeline consists of two parts: 1. data preprocessing and 2. training countermeasure models. In the data preprocessing step, we employ a pre-trained speech enhancement model to amplify the artifacts in the utterances, which involves three steps: 1. noise addition, 2. noise extraction, and 3. noise amplification. Then, we train the countermeasure model using the amplified utterances. In this section, we explain the three preprocessing steps in detail.

\subsection{Noise Addition}
In the first step, we introduce noise into the utterance at a target signal-to-noise ratio (SNR) to simulate noisy speech, as defined by the following equation:
\begin{align}
    y &= x + \sqrt{\frac{||x||^2}{||n||^2 \times 10^{\frac{\text{SNR}}{10}}}} \cdot n
    % r_\text{snr} &= \sqrt{\frac{||x||^2}{||n||^2 \times 10^{\frac{\text{SNR}}{10}}}}
\end{align}
where $x \in \mathbb{R}^t$ and $n \in \mathbb{R}^t$ represent the raw input utterance and noise signal, respectively, with $t$ denoting the utterance duration (in samples). The operator $||\cdot||$ denotes the L2 norm. The noise signal $n$ can be any type of noise, such as white noise, brown noise, or other environmental noise. The resulting noisy speech is then fed into the speech enhancement model.

\subsection{Noise Extraction}
We assume that the speech enhancement model not only removes noise $n$, but also eliminates artifacts presented in spoofed utterances. We claim that the removed artifacts contain valuable information to help the countermeasure detect spoofed utterances. To leverage this, we introduce a method to extract artifact-contained noise from the output of the speech enhancement model.

% New Version
In speech enhancement, we typically assume that clean speech and noise are orthogonal. However, \cite{si_sdr} demonstrated that directly subtracting clean speech from the estimated clean speech does not guarantee orthogonality. This occurs because the estimated clean speech often has a different scale compared to the target clean speech, leading to a residual clean speech component in the estimated noise. Following the same assumption, we adopt a similar approach as follows:
% inaccurate signal-to-noise ratio of the speech enhancement output due to residual clean speech in the estimated noise. Therefore, \cite{si_sdr} proposed projecting the target onto the output to ensure they are on the same scale. 
\begin{align}
    \hat{a} = x - \frac{x\cdot \hat{x}}{||\hat{x}||^2} \hat{x}
\end{align}
% \begin{equation}
%     N = X - proj_\text{raw} \hat{X} = X - \frac{(X \cdot \hat{X})}{||\hat{X}||^2} \cdot \hat{X}
% \end{equation}
where $\hat{x}$ is the output of the speech enhancement model, $x$ is the raw input utterance, $\frac{x \cdot \hat{x}}{||\hat{x}||^2}$ is referred to as the \textit{projection weight}, and $\hat{a}$ is the estimated entangled noise and artifact. We hypothesize that the artifact components extracted from spoofed utterances should have a higher magnitude than those from bonafide utterances, as bonafide speech does not contain artifacts. Therefore, speech enhancement acts as an artifact extractor for spoofed speech while functioning as a regular noise reducer for bonafide speech.

\subsection{Noise Amplification}
In this step, we amplify the extracted noise by adding it back into the raw input utterance $x$, as shown in the equation below:
\begin{align}
    \tilde{x} = x + \alpha \hat{a}
\end{align}
where $\alpha$ is a hyperparameter that controls the degree of noise amplification in the utterance. If the input $x$ is a spoofed utterance, entangled noise and artifacts will be amplified and added back to the speech for all $\alpha > 0$, making it easier for the countermeasure to detect them. In contrast, only the noise will be amplified and added back to bona fide speech. As a result, by amplifying the artifacts, the countermeasure model should be better able to distinguish between bona fide and spoofed utterances. We selected the best alpha based on our grid search parameters, in which $\alpha=1.4$ for LCNN and RawNet2, and $\alpha=0.6$ for AASIST.

\section{Experimental Setup}

\subsection{Datasets}

\subsubsection{VoiceBank + DEMAND}
Speech enhancement models were pre-trained on the VoiceBank + DEMAND dataset, consisting of 30 speakers from the VoiceBank Corpus and eight recorded noise samples from DEMAND. The training and validation sets contain 11,572 utterances corrupted by eight recorded noise samples from DEMAND and two artificially generated noise samples (babble and speech-shaped) at SNR levels of 0, 5, 10, and 15 dB, while the testing set contains 824 utterances, each contaminated with five unseen recorded noise samples from DEMAND, at SNR levels of 2.5, 7.5, 12.5, and 17.5 dB. All speech data were sampled at 16 kHz. The training procedure of the speech enhancement models follows Thunder \cite{trachu24_interspeech}.

\subsubsection{ASVspoof}
We benchmarked the performance of our proposed method on both the ASVspoof2019 \cite{asvspoof2019} and ASVspoof2021 \cite{asvspoof2021} logical access (LA) datasets. The ASVspoof2019LA dataset consists of three subsets: train, development, and evaluation. Spoofed utterances are generated using 19 spoofing attack algorithms (A01-A19), including speech synthesis and voice conversion. 

In the training and development sets, spoofed utterances are generated from six different algorithms, while the evaluation set contains spoofed utterances generated by 13 different algorithms. The ASVspoof2021LA dataset, on the other hand, employs the same training and development sets as the ASVspoof2019LA dataset and only introduces a new evaluation set. In this new evaluation set, six different codecs were applied to both bona fide and spoofed utterances from the ASVspoof2021LA evaluation set. We computed the signal-to-noise ratio (SNR) using WADA-SNR \cite{kim08e_interspeech}, and the average SNR for ASVspoof2019 and ASVspoof2021 eval sets is 67.82 and 73.11, which are considered clean speech. %Regarding ASVspoof2024, we chose to exclude it from our evaluation as the test set is not publicly available.

\subsection{Training Details and Evaluation Measures}
% how long for the utterace
\subsubsection{Speech Enhancement}
We used the official implementation of Thunder\footnote{https://github.com/SLSCU/thunder-speech-enhancement} to pre-train the Thunder, NCSNPP, and GaGNet models on the VoiceBank+DEMAND dataset. All models were trained for 100 epochs on an A100 GPU using the Adam optimizer, a learning rate of $2 \times 10^{-5}$, and a batch size of 8. For other hyperparameters, we followed the default settings from their respective papers. We then selected the epoch with the lowest validation loss as the pretrained model. No augmentation was used during training.

\subsubsection{Countermeasures}
Our code was mainly based on the ASVspoof2021 baseline\footnote{https://github.com/asvspoof-challenge/2021/tree/main/LA/Baseline-RawNet2}. Anti-spoofing countermeasure models were trained for 30 epochs on an A100 GPU, using the Adam optimizer, with a learning rate of $1e-4$ that decayed by $0.9$ after 10 epochs and a batch size of 32. White noise were used with the default 0 dB SNR in the noise addition process. The utterances were randomly cropped into 4 seconds for both training and evaluation. We used the tandem detection cost function (t-DCF) \cite{kinnunen2020tandem} and equal error rate (EER) as primary metrics, following the ASVspoof2021 challenge \cite{asvspoof2021}. We did not perform any augmentation method in our pipeline.

% trim to 4 sec and pad (using wrap) to 4 sec, 

% Explain Metrics EER and min tdcf

\section{Results and Discussion}
\begin{table}[t]
    \centering
        \caption{Performance of different speech enhancement and countermeasure models on the ASVspoof2019 and ASVspoof2021 datasets. ``SE'' and ``CM'' denote the speech enhancement and countermeasure models, respectively. The best-performing model in each section is bolded.}
    \begin{tabular}{cccccc}
    
        \toprule
        \multicolumn{2}{c}{\textbf{Model}} & \multicolumn{2}{c}{\textbf{2019}} & \multicolumn{2}{c}{\textbf{2021}}\\
        \cmidrule(lr){1-2} \cmidrule(lr){3-4} \cmidrule(lr){5-6}
        SE & CM & t-DCF & EER & t-DCF & EER \\
        \midrule
        - & LCNN & 0.136 & 6.20 & 0.341 & 7.21 \\
        Thunder & LCNN & \textbf{0.102} & \textbf{4.46} & \textbf{0.310} & 5.56 \\
        GaGNet & LCNN & 0.117 & 4.84 & 0.322 & 5.82 \\
        NCSNPP & LCNN & 0.106 & 4.70 & 0.312 & \textbf{5.31} \\
        \midrule
        - & AASIST & 0.048 & 1.81 & 0.370 & 6.70 \\
        Thunder & AASIST & 0.043 & 1.55 & \textbf{0.339} & \textbf{5.87} \\
        GaGNet & AASIST & \textbf{0.035} & \textbf{1.24} & 0.363 & 6.55 \\
        NCSNPP & AASIST & 0.057 & 2.07 & 0.358 & 6.57 \\
        \midrule
        - & RawNet2 & 0.110 & 4.83 & 0.385 & 8.10 \\
        Thunder & RawNet2 & \textbf{0.062} & 2.68 & 0.349 & 6.88 \\
        GaGNet & RawNet2 & 0.065 &\textbf{2.65} & 0.359 & 7.09 \\
        NCSNPP & RawNet2 & 0.078 & 3.33 & \textbf{0.341} & \textbf{6.67} \\
        \bottomrule
    \end{tabular}
    \vspace{2mm}

    \label{tab:main_result}
    \vspace{-7mm}
\end{table}
\subsection{Main Results}
We evaluated the pipeline with multiple combinations of speech enhancement (SE) and countermeasure (CM) models to demonstrate its robustness across different architectures. Both discriminative and generative speech enhancement models work well with CNN-based and GNN-based countermeasures. Table \ref{tab:main_result} shows that our method is effective across various speech enhancement and countermeasure models, compared to using the countermeasure model alone. Moreover, it generalizes well to both ASVspoof2019 and ASVspoof2021. Performance is relatively improved by up to 45\% in EER and 43\% in t-DCF on ASVspoof2019, and when evaluated on ASVspoof2021, the pipeline demonstrates a relative performance gain of up to 26\% in EER and 11\% in t-DCF.

% In the following analysis, we primarily use Thunder and RawNet2 for the ablation study. We choose this combination because it achieves the lowest average EER across ASVspoofed 2019 and 2021.

% An overview of our pipeline is presented in Figure \ref{fig:pipeline}. Our pipeline is model-agnostic, as we evaluate it with different speech enhancement (SE) and countermeasure (CM) models, including both generative and discriminative SE models, as well as frequency-based and time-based CMs. This ensures the robustness of our approach across different architectures. Our evaluation shows that both discriminative and generative speech enhancement models work well with various countermeasure types.
% Performance can be relatively improved up to xx in EER and xx in min dcf on ASVspoofed2019. When evaluated with ASVspoofed2021, the pipeline gives a relative performance gain of up to xx in EER and xx in min dcf.
\subsection{Impact of Generated Noise}

The objective of adding noise is to encourage the speech enhancement model to be more aggressive in removing noise and artifacts, as we observed greater SNR improvements when fed low-SNR inputs. We conducted an experiment on ASVspoof2021 with Thunder, comparing systems with and without the noise addition step. Table \ref{tab:impact_noise} shows that adding white noise before speech enhancement helps the model extract artifacts more effectively, achieving better results than without noise addition. Speech enhancement typically generates artifacts that degrade the performance of downstream tasks, as observed in ASR tasks \cite{iwamoto22_interspeech}. Without the noise from the noise addition process, the extracted noise would primarily consist of speech enhancement artifacts, making it difficult for the countermeasure model to distinguish between spoofed and bona fide utterances. 

% We observed that the model becomes more aggressive at lower input SNRs, which helps it extract more artifacts---showing gains of 10~dB and 5~dB at input SNRs of 0~dB and 10~dB, respectively.

% We also observed that the speech enhancement model tends to be more aggressive at lower input SNR levels, showing improvements of 10 dB and 5 dB when the input SNR is 0 dB and 10 dB, respectively. We believe this aggressiveness may help in extracting artifacts.

% Conversely, when noise is added, the speech enhancement model extracts TTS or voice conversion artifacts from spoofed utterances, enabling the countermeasure model to better differentiate them from bona fide utterances.

% In contrast, without noise addition, performance deteriorates, which we assume happens because the energy of noise is not sufficient for the model to extract artifact-contained noise. As a result, amplifying the extracted noise may conceal important features rather than enhance the artifact.

% In contrast, without noise addition, performance deteriorates, which we assume happens because the speech enhancement model would also produce an artifact, according to \cite{iwamoto22_interspeech}.

\begin{table}[h]
    \vspace{-3.5pt}
    \centering
        \caption{Performance of different countermeasures in the absence of white noise on the ASVspoof2021 dataset. The noise addition step is crucial for speech enhancement to remove artifacts from the utterance.}
    \begin{tabular}{cccc}
        \toprule
        CM & Noise Addition & t-DCF & EER   \\
        \midrule
        LCNN & Without noise & 0.352 & 8.67 \\
        LCNN & With noise & \textbf{0.310} & \textbf{5.56} \\
        \midrule
        AASIST & Without noise & 0.356 & 7.09 \\
        AASIST & With noise & \textbf{0.339} & \textbf{5.87} \\
        \midrule
        RawNet2 & Without noise & 0.478 & 10.46 \\
        RawNet2 & With noise & \textbf{0.349} & \textbf{6.88}\\
        \bottomrule
    \end{tabular}
    \vspace{2mm}
    \label{tab:impact_noise}
    \vspace{-8mm}
\end{table}

\subsection{Influence of Generated Noise Types}
We evaluated our method on ASVspoof2021 using three types of noise: white, violet, and pink noise, each with distinct characteristics. White noise has equal power across all frequency bands, while violet noise has greater intensity at higher frequencies, and pink noise has higher intensity at lower frequencies. Using different types of noise helps the speech enhancement model detect artifacts across various frequency bands. As shown in Table \ref{tab:noise_type}, the best performance is achieved when using white noise, suggesting that the artifacts are present across all frequency bands. Furthermore, the model performs worse with both pink noise and violet noise compared to white noise, possibly due to challenges in detecting artifacts in certain frequency bands. 
Moreover, as suggested in \cite{10446798}, different vocoders create different kinds of artifacts. Our analysis revealed that different spoofing algorithms generate artifacts in different frequency bands. For example, A08 is more easily detected with violet noise, while A18 is better detected with pink noise.

% To further investigate this, we applied low-pass and high-pass filters to the extracted noise from the white noise experiment. The results remained consistent, with the low-pass filter yielding better performance than the high-pass filter.
% Moreover, our analysis supports the findings in \cite{10446798}, which state that different vocoders generate different types of artifacts. We found that some attacks, such as A08, are more easily detected with violet noise, while others, like A18, are better detected with pink noise.

% Furthermore, the model performs better with pink noise than with violet noise, indicating that the speech enhancement model extracts low-frequency noise better than high-frequency noise when the added noise has higher density in low-frequency bands.
% indicating that artifacts are predominantly concentrated in the lower frequency bands, where pink noise has a higher power density compared to violet noise.

\begin{table}[h]
    \vspace{-3.5pt}
    \centering
        \caption{Performance across different noise types. White noise yields the best results. Pink noise outperforms violet noise, indicating artifacts are more prominent in lower frequencies.}
    \begin{tabular}{ccccc}
        \toprule
        SE & CM & Noise & t-DCF & EER   \\
        \midrule
        - & RawNet2 & - & 0.385 & 8.10 \\
        Thunder & RawNet2 & Violet & 0.358 & 7.53\\
        Thunder & RawNet2 & Pink & 0.351 & 7.01 \\
        Thunder & RawNet2 & White & \textbf{0.349} & \textbf{6.88} \\
        \bottomrule
    \end{tabular}
    \vspace{2mm}

    \label{tab:noise_type}
    \vspace{-8mm}
\end{table}

\section{Ablation Studies}
\subsection{Effect of Using Projection Weight}
% introduce the concept of speech enhancement model 
We validated our projection approach in the noise extraction step on the ASVspoof 2021 dataset using Thunder and RawNet2. We compared the performance of scaling the output of the speech enhancement model with projection weights versus using no scaling. Without scaling, we expect that the extracted artifacts will retain residual clean speech components, making them less distinct and hindering the countermeasure model's ability to detect them. As shown in Table \ref{tab:effect_projection}, our proposed projection method in the noise extraction layer improves performance by approximately 8\% compared to the naive approach. This result supports our hypothesis that clean speech components remain in the extracted noise from speech enhancement. Directly using the extracted noise from the naive approach amplifies and reintroduces these clean speech components into the input of the countermeasure model, leading to inferior performance compared to the projection-based method.

\begin{table}[h]
    \vspace{-3.5pt}
    \centering
        \caption{Performance across different noise extraction methods. Using projection outperforms the naive approach as it extracts only the artifact information.}
    \begin{tabular}{ccccc}
        \toprule
        SE & CM & Noise Extraction & t-DCF & EER   \\
        \midrule
        - & RawNet2 & - & 0.385 & 8.10 \\
        Thunder & RawNet2  & without Projection & 0.378 & 7.48 \\
        Thunder& RawNet2 & with Projection & \textbf{0.349} & \textbf{6.88} \\
        \bottomrule
    \end{tabular}
    \vspace{2mm}

    \label{tab:effect_projection}
    \vspace{-8mm}
\end{table}

\subsection{Effect of the SNR of the Generated Noise}

The SNR of the added noise determines the energy ratio between the speech component and the artifact component, affecting  the noise extraction and amplification processes. We performed the experiment with different SNR levels and evaluated the performance on both ASVspoof2019 and ASVspoof2021, using Thunder and RawNet2 as the speech enhancement and the countermeasure model.

As shown in Table \ref{tab:snr_effect}, adding too much or too little noise can be detrimental to the framework's performance. In the former case, excessive noise may obscure the presence of the artifacts as entangled noise and artifacts are amplified by the same $\alpha$ during the noise amplification process, maintaining the energy ratio. When the SNR between noise and artifacts is too high, the model may interpret it as a typical noisy speech. In the latter case, insufficient noise may prevent the speech enhancement model from detecting artifacts in certain frequency bands. Our results indicate that using $SNR = 0$ is the most optimal SNR, which can relatively improve up to 33.82\% in EER compared with $SNR= 10$.

\begin{table}[h]
    \vspace{-3.5pt}
    \centering
        \caption{Effect of SNR levels in the noise addition step.}
    \begin{tabular}{ccccc}
        \toprule
         & \multicolumn{2}{c}{\textbf{2019}} & \multicolumn{2}{c}{\textbf{2021}}\\
         \cmidrule(lr){2-3} \cmidrule(lr){4-5}
        SNR & t-DCF & EER & t-DCF & EER \\
        \midrule
        -5 & 0.080 & 3.75 & 0.363 & 7.28 \\
        0 & \textbf{0.062} & \textbf{2.68} & \textbf{0.349} & \textbf{6.88} \\
        5 & 0.066 & 2.86 & 0.379 & 7.41 \\
        10 & 0.093 & 4.05 & 0.386 & 8.13 \\
        \bottomrule
    \end{tabular}
    \vspace{2mm}

    \label{tab:snr_effect}
    \vspace{-8mm}
\end{table}

% \begin{figure}[ht]
%     \centering
%     \includegraphics[scale=0.5]{figures/ablation_alpha.pdf}
%     \caption{Effect on difference alpha on different SNR.}
%     \label{fig:alpha_ablation}
% \end{figure}

% The Figure \ref{fig:alpha_ablation} shows that the higher the SNR for noisy speech, the worse the result, which is consistent with Table \ref{tab:impact_noise}. Moreover, we also show that different $\alpha$ can lead to different performance. The lower the alpha $\alpha$ is, the less we amplify the artifacts. However, when the $\alpha$ is too high, the noise-amplified utterances behave like a noisy speech for both bona fide and spoofed, making them harder to be classified.

%  utt -> +noise(1) -> se -> extract(2) -> amp ->cm
% 1 -> frequency domain -> noises (3)
% 2 -> frequency domain -> extracted artifacts with noises (4)
% (4) - (3) = artifacts for spoofed utt

%  utt(5)-> +noise(1) -> se -> extract(2) -> amp ->cm
% 1 -> frequency domain -> noises (3)
% 2 -> frequency domain -> extracted artifacts with noises (4)
% 5 -> se -> clean speech -> extract noise -> frequency domain (6)
% (4)-(3)-(6) = artifacts for spoofed utt

% do this with both bona fide and spoofed, and compare them together
% \begin{table}[]
%     \centering
%     \begin{tabular}{cccc}
%         \toprule
%         set & bona fide & spoofed & pooled   \\
%         \midrule
%         train & 97.30 & 87.12 & 88.15 \\
%         dev & 98.92 & 87.68 & 88.83 \\
%         eval2019 & 99.18 & 64.21 & 67.82 \\
%         eval2021 & 99.74 & 70.15 & 73.11 \\
%         \bottomrule
%     \end{tabular}
%     \caption{wada snr}
% \end{table}

\section{Conclusion}
We propose a pipeline for amplifying artifacts in speech utterances using speech enhancement. The pipeline comprises three steps. First, some noise is added to the raw utterance to simulate noisy speech, prompting the speech enhancement model to remove artifacts. Next, we feed the noisy speech into the speech enhancement model and extract the noise from the output before amplifying the noise back into the clean speech. Finally, we use this noise-amplified data to train the countermeasure model. Our method is compatible with various speech enhancement and countermeasure models and improves countermeasure performance on both the ASVspoof2019 and ASVspoof2021 datasets.

\bibliographystyle{IEEEtran}
\bibliography{mybib}

\end{document}